\documentstyle[12pt,aps,manuscript,eqsecnum]{revtex}
\newcommand{\ia}{in\-ter\-ac\-tion}
\newcommand{\as}{an\-ti\-sym\-met\-ric}

\newcommand{\ds}{den\-si\-ty of states}
\newcommand{\dss}{den\-si\-ties of states}
\newcommand{\me}{ma\-trix ele\-ment}
\newcommand{\sw}{sprea\-ding width}
\newcommand{\swd}{\Gamma^{\downarrow}}
\newcommand{\ee}{\end{equation}}
\newcommand{\be}[1]{\begin{equation} \label{#1}}
\newcommand{\ba}[1]{\begin{eqnarray} \label{#1}}
\newcommand{\ea}{\end{eqnarray}}
\newcommand{\nn}{\nonumber}
\newcommand{\ket}[1]{|{#1}\rangle}
\newcommand{\bra}[1]{\langle{#1}|}

\newcommand{\Bracket}[4]{\langle \, ^{#1}_{#2}| \, ^{#3}_{#4}\rangle}
\newcommand{\K}[2]{| \, ^{#1}_{#2} \rangle}
\newcommand{\B}[2]{\langle \, ^{#1}_{#2} |}
\newcommand{\V}[3]{ _{\phantom{()}{#1}} \hspace{-0.05em}{\sf V}_{({#2},{#3})}}

\newcommand{\St}[5] { ^{\phantom{()}{#1}}_{\phantom{()}} \hspace{-0.2em}
                    S^{({#2},{#3})}_{({#4},{#5})}}
\newcommand{\Sv}[3] { ^{\phantom{()}{#1}} \hspace{-0.2em}
                    S^{({#2},{#3})}}

\newcommand{\D}[3]{ \rho^{\phantom{()}}_{{#1},{#2}}({#3}) }
\newcommand{\DE}[3]{ \rho^{({\rm Er})}_{{#1},{#2}}({#3}) }

\newcommand{\es}[1]{\sum_{ \{ {#1} \}}} 
\newcommand{\des}[2]{\sum_{ \{ {#1}{#2} \}}} 
\newcommand{\mes}[3]{\sum_{ \{ {#1};\,{#2}\,;{#3} \}}} 
\newcommand{\ves}[2]{\sum_{ \{ {#1};\,{#2} \}}} 

\begin{document}
\draft
\title{Propagation of a $K$-body force into $A$-body space}

\author{M. Granzow, H. L. Harney, and H. Kalka$^*$}

\address{Max--Planck--Institut f\"{u}r Kernphysik,
69029 Heidelberg, FRG\\
$^*$Institut f\"{u}r Theoretische Physik der Technischen
Universit\"{a}t, 01062 Dresden, FRG}
\date{submitted to {\sl Phys.~Rev.~C}}

\maketitle

\begin{abstract}
The calculation of the spreading width of a compound nuclear reaction
caused by a symmetry breaking $K$-body force acting in an $A$-body
system ($K \ll A$, usually $K=2$) involves the determination of the
local average square matrix element in $A$-body space. This problem is
reduced to finding the global mean square matrix element $v^2$ in
$K$-body space. The result is a compact formula for the spreading
width which contains ${v^2}$ as an input. Our method is based on the
dilute gas approximation for excitons close to the Fermi edge. The
relative strength of the contributions of operators with different
exciton rank as well as the connection between the energy dependence
of the spreading width and the body rank of the underlying interaction
are established.
\end{abstract}


\section{Introduction}
The compound nucleus (CN) has received considerable attention --- both
theoretical and experimental --- in the recent years because
of its seeming ability to enhance the effects of the weak interaction
to a few percent (see for example~\cite{Fr} and references
therein). In the present study we consider the
breaking of symmetries caused by $K$-body forces in the compound
process in general. The adequate quantity to characterize symmetry
nonconservation in many-body systems is the \sw \ due to the
underlying interaction\cite{RH}. Although in most applications a
two--body force is used, we treat the case of arbitrary $K$ since it
will allow for some insight into the energy dependence of the
\sw . This provides a generalization of the
results presented in~\cite{yo} and the communication of some details
omitted therein.

In the statistical theory of CN reactions the \me s of the interaction
are assumed to show the characteristics of the Gaussian orthogonal
ensemble (GOE). The crucial point is that statistical assumptions of
this kind can only be made about the defining \me s, i.e. the \me s in
$K$-body space. However, the quantity of physical interest is the \me
\ in $A$-body space. The connection between the properties of the
$K$-body \me s and those in $A$-body space is called the propagation
of the defining \me s~\cite{Wong}.  In the present article a solution
for this problem is offered which consists of the transition to the
exciton picture (with the accompanying simplification of the basis
states and complication in the description of the \ia), the
propagation into the subspace of fixed exciton number, and the
averaging over subspaces, which implies the return to the body
picture. This procedure is necessary because our formalism
makes use of the dilute gas approximation (DGA) which --- in contrast
to the body picture --- is very good in the exciton representation of
the system. In the final sections then, the general expression, which
involves a convolution of partial and total \dss , is evaluated by
inserting well known analytical formulae for the level densities, the
results are discussed, and limitations of our approach are indicated.

\section{Concepts}\label{con}
\subsection{Spaces and bases}
The physical quantity we are interested in is the \sw
\be{spreadw}
\swd(E) = 2\pi \ll {\sf V}^2 \gg \rho (E) \,\, .
\ee
Here, $\ll{\sf V}^2 \gg$ is the mean square \me \ of the $K$-body \ia
\ ${\sf V}$, and $\rho$ is the level density of the system. The \me s
are calculated in a basis of eigenstates to those parts of the
Hamiltonian that dominate the behaviour of the system. The \sw \ due
to additional, symmetry breaking \ia s then measures the extent of
symmetry breaking in the system.  This may include the breaking of the
independent particle structure, isospin symmetry or parity by the
residual strong, the electromagnetic or weak \ia , respectively. In
order to properly treat the variation of $\swd$ with energy, the
average $ \ll {\sf V}^2 \gg$ is limited to states in the neighbourhood
of some given excitation energy $E$. In fact, part of the present work
will consist of the calculation of the strength function
\be{locstrA}
S(E',E) \ = \ \ll {\sf V}^2 \gg \rho (E')\rho (E)
\ee
which implies the average over squared \me s between configurations
close to $E$ and configurations close to $E'$. Once the basis has been
specified in detail, this {\sl local} average will be defined in
section~\ref{proc} .  The basis we work with is built up from the
single particle states $\ket{\lambda}$ that satisfy the canonical
Hartree--Fock equations
\be{kHF} ({\sf u + u}_{\rm HF} ) \ket{\lambda}
=:{\sf h}_0 \ket{\lambda} = \varepsilon_{\lambda} \ket{\lambda} \,\, ,
\lambda=1 \ldots D \, .
\ee
Here, $D$ is the dimension of the one--body space spanned by the
discrete set of bound states, ${\sf u}$ is the
one-body kinetic energy operator, and ${\sf u}_{\rm HF}$ is the
Hartree-Fock mean field operator constructed from the strong
nucleon-nucleon \ia , which does not include the symmetry breaking
\ia \ ${\sf V}$ for which the \sw \ shall be calculated.  Let us
introduce the creation operators $ {\sf b}^\dagger_{\lambda}$ and the
physical vacuum $ \ket {\phantom{0}}$ so that

\be {EKZ} \ket {\lambda} = {\sf b}^\dagger_{\lambda} \ket
{\phantom{0}} \,\, .
\ee
These operators fulfill the fundamental
anticommutation relation
\be {AVR} \{{\sf b}_{\lambda} \, , \, {\sf
b}^\dagger_{\lambda'}\} = \delta_{\lambda \lambda' } \,\, .
\ee
A
basis of $A$-body states is then given by all possible $A$-fold
applications of creation operators:
\be{Abodystate} \ket{\Lambda}
\equiv \ket{\lambda_1 \ldots \lambda_A} \equiv {\sf
b}^\dagger_{\lambda_A} \ldots {\sf
b}^\dagger_{\lambda_1}\ket{\phantom{0}} \,\, ,
\ee
with the condition
\be{basiscond}
\lambda_1 < \lambda_2 < \ldots <
\lambda_A.
\ee
The energy ${\cal E}_\Lambda$ of this many-body state is
\be{basisener}
{\cal E}_\Lambda = \sum_{i=1}^A \varepsilon_{\lambda_i} \,\, .
\ee
The energy of the ground state $\ket{1 \ldots A}$ is $
\sum_{i=1}^A \varepsilon_i$ and will be denoted ${\cal E}_A$.  The
states~(\ref{Abodystate}) span the whole $A$-body space of dimension
${D \choose A }$, but for low energy considerations it is convenient
to split the $A$-body space into subspaces of increasing complexity
and energy and to restrict the description of the system to the
subspaces in the energy range of interest. Technically this idea is
realized by the transition to the exciton picture. For moderate
excitation energies the state of the system does not strongly differ
from the ground state and there will be only a few states occupied
above and unoccupied below the Fermi energy ${\varepsilon}_{A}$.
This will make it possible --- in
section~\ref{PS} --- to introduce the so-called dilute gas
approximation and treat the present problem in closed form.
Let us therefore introduce the exciton creation
operators~\cite{RS,FW,BM} ${\sf a}^\dagger_{\lambda}$ according
to~\cite{time}
\ba{exop}
{\sf a}^\dagger_{\lambda} \equiv {\sf
b}^\dagger_{\lambda} \hspace{1cm} &{\rm for}& \,\, \lambda>C \nn \\
{\sf a}^\dagger_{\lambda} \equiv {\sf b}_{\lambda} \hspace{1cm} &{\rm
for}& \,\, \lambda \leq C
\ea
acting on the exciton vacuum
\be{exopvac} \ket{0} \equiv \ket{1 \ldots C} \,\, .
\ee
One then has
${\sf a}_{\lambda} \ket{0} = 0$ for all $\lambda$ and the
anticommutation relation~(\ref{AVR}) holds for the operators ${\sf
a}$, too. Briefly: the excitons are fermions. The single body space is
split into two subspaces, the single particle space with dimension $d_p
\equiv D-C$ and the single hole space with dimension $d_h \equiv
C$. The exciton vacuum $ \ket{0}$ --- the core --- need not be identical
to the ground state of the $A$-body system under consideration,
i.e. at present we do not fix
\be{Delta}
\Delta \equiv A-C \,\, .
\ee
The energy of the vacuum is called \be{vacener} {\cal E}_C \equiv
\sum_{\lambda=1}^C \varepsilon_{\lambda} \,\, .  \ee Every configuration
$\ket{\Lambda}$ can then be characterized by a vector \be{P} P \equiv
(r_1 , \ldots ,r_p) \hspace{1cm} ,\hspace{1cm} r_i > C \ee of its
particle configurations and by a vector
\be{H}
H \equiv (\alpha_1 ,
\ldots ,\alpha_h) \hspace{1cm} , \hspace{1cm} \alpha_i \leq C
\ee
of
its hole configurations. Every independent particle configuration
$\ket{\Lambda}$ can be expressed in the form
\be{PH}
\K PH \equiv {\sf a}^\dagger_{r_p}
\ldots {\sf a}^\dagger_{r_1}{\sf a}^\dagger_{\alpha_1} \ldots {\sf
a}^\dagger_{\alpha_h} \ket 0 \,\, .
\ee
The combination of operators
appearing in this equation will also be written as
\be{APH}
{\sf a}^\dagger_{r_p} \ldots {\sf a}^\dagger_{r_1}{\sf
a}^\dagger_{\alpha_1} \ldots {\sf a}^\dagger_{\alpha_h} \equiv ({\sf
A}^P_H)^{ \dagger}  \,\, .
\ee
As usual we introduce the energies
\be{enerp}
\epsilon_{r_i} \equiv {\varepsilon}_{r_i} - \varepsilon_C
\ee
of the particle configurations relative to the core level as well as
the energies of
the hole configurations
\be{enerh} \epsilon_{\alpha_i} \equiv {\varepsilon}_C
- {\varepsilon}_{\alpha_i} \,\, .
\ee
The energy of the state~(\ref{PH}) then is
\be{ener}
{\cal E}^P_H \equiv \sum_{i=1}^h
\epsilon_{\alpha_i} + \sum_{i=1}^p \epsilon_{r_i} + {\cal E}_C +
{\varepsilon}_{C} \Delta \equiv \epsilon^P_H + {\cal E}_C +
{\varepsilon}_{C} \Delta \,\, .
\ee
The symbol $ \epsilon^P_H $ has been introduced to simplify the
notation. In the present paper the letters $r,s,t,u$ or $v$
will be used for particle states and $\alpha,\beta,\gamma,\delta$ or
$\epsilon$ for hole states; configurations of the physical
constituents of the system --- called {\sl bodies} --- are labelled
$\lambda,\mu,\nu,\rho$ or $\sigma$.  If one chooses $C<A$, the
ground state of the system has $\Delta$ particles and no holes,
otherwise it has no particles and $-\Delta$ holes. Generally one has
\be{pminush}
p-h =\Delta
\ee
and
\be{phmin}
p \geq p_{\rm min} = {\rm max}
(\Delta,0) \,\, , \hspace{0.7cm} h \geq {\rm max} (-\Delta,0) \,\, .
\ee
The maximum number of particles is $A$ and that of holes
$C=A-\Delta$. Hence, the exciton picture provides a decomposition of
the $A$-body space into mutually orthogonal subspaces ${\cal U}_{p}$,
$p=p_{\rm min}, \ldots, A$. The dimension of the  subspace ${\cal U}_{p}$
is $d({\cal U}_p)={d_h \choose h}{d_p \choose p}$.  We use the
following notation for many-exciton states: A state $ \K{P'}{H'}$
contains the same numbers of particles and holes as $\K PH$ on
possibly different single exciton states.  A ket $\K TL$ on the other
hand differs from $\K PH$ with respect to exciton number as well as
single exciton states.

\subsection{Conventions of summation}
The spectral average over the squared \me s of ${\sf V}$ implies sums
over the basis states~(\ref{PH}). To avoid double counting of the
basis states a definite order of the indices $r$ and $\alpha$ in
eq.~(\ref{PH}) must be observed. We therefore introduce the notation
\be{eingsum}
 \es H
\equiv
 \sum_{\alpha_1< \ldots <\alpha_h} \,\, .
\ee
Here, the $\alpha_i$ with $i=1, \ldots , h$ run over the
configurations $1 \leq \alpha_i \leq C$. The sum is called restricted
because of the restriction $\alpha_1< \ldots <\alpha_h$ imposed on the
permitted terms. Hence, the sum has ${d_h \choose h}$ terms.
In contrast the sum
\be{uneingsum}
 \sum_H
\equiv
 \sum_{\alpha_1 \ldots \alpha_h}
\ee
is called unrestricted and has $(d_h)^h$ terms. Corresponding
conventions are used for the set $P$ of particle configurations.
For later purposes we compress the notation of~(\ref{eingsum})
further:
\be {multeingsum}
 \mes{HP} \ldots {LT}
\equiv
 \des HP \ldots \des LT \,\, ,
\ee
which means that the ordering has to be observed only within each
group of indices separated by semicolons.
In the following use will be made of the identity
\be{iden}
 \es H \B {T}{L} {\sf V} \K PH^2 =
         \frac{1}{h!} \sum_H \B {T}{L} {\sf V} \K PH^2 \,\, ,
\ee
which holds because the {\sl squared} \me \ is completely symmetric with
respect to the $h$ indices in $H$, but vanishes if any two of them coincide.

\subsection{Interactions for bodies and excitons}
Now that the basis we are going to work with is specified, we turn to
those parts of the \ia \ that have not been considered in its
determination and consequently cause transitions between basis
states. Apart from the residual strong \ia \ this includes the
electromagnetic, weak, and other possible forces. In occupation number
formalism, a $K$-body operator
\be{operator}
{\sf V}   =  \sum_{i_1< \ldots < i_K = 1}^A {\sf v}(x_{i_1} \ldots
             x_{i_K})
\ee
has the form
\be{opbzd}
{\sf V}     =   \ves{\mu_1, \ldots, \mu_K}{\nu_1, \ldots, \nu_K}
                \bra{\mu_1 \ldots \mu_K}
                \tilde{{\sf v}}
                \ket{\nu_1 \ldots \nu_K}
                {\sf b}^\dagger_{\mu_K} \ldots
                {\sf b}^\dagger_{\mu_1}
                {\sf b}_{\nu_1} \ldots
                {\sf b}_{\nu_K} \,\, ,
\ee
with the totally \as \ \me \ $ \bra{\mu_1 \ldots \mu_K} \tilde{{\sf
v}} \ket{\nu_1 \ldots \nu_K} $.  How is this representation affected
by the transition to the exciton picture? The range of summation of
every index is split into two parts, body operators are replaced by
exciton operators according to eq.~(\ref{exop}), and the resulting
terms are brought into normal order and grouped according to
particle-hole structure. For $K=2$, the result is given in
refs.~\cite{RS,BM,HMF} and is reproduced in Tab.~I. The \ia
\ is the sum of the fourteen terms $\V kaq$ listed there together with
their Feynman diagrams~\cite{HMF}. The diagrams facilitate the
visualization of the systematics in and the generalization of the
contents of Tab.~I, see Fig.~\ref{2biaep} . The exciton
operators can be classified by the three numbers $k,a$, and $q$. The
rank $k$ of $\V kaq$ is half the number of external lines in the
corresponding diagram. One finds $0 \leq k \leq K$. The parameter $a$
is the number of particle-hole pairs created by $\V kaq$: the \ia \
does not conserve the number of excitons; the conservation of the
number of physical bodies, however, requires that the exciton number
changes by particle-hole pairs. The range of $a$ obviously is $-k \leq
a \leq k$. Finally, $q$ is the number of upward arrows in the
diagram. One recognizes that $q$ changes in steps of two, since for
fixed $a$ every additional particle before the interaction leads to an
additional particle after the \ia . The range of $q$ is found to be
$|a| \leq q \leq 2k-|a|$. Hence,
\be{sumVkaq}
          {\sf V} = \sum_{k=0}^K \sum_{a=-k}^k \sum_{
          {q=|a| \atop \Delta q=2} }^{2k-|a|} \, \V kaq.
\ee
On the table, the operators with $k<2$ have
contracted hole lines that represent the interaction of the excitons
with the nuclear core.
It is easily seen (and holds for arbitrary $K$ as well) that the
number of particle and hole lines before and after the interaction is
given by $N_- \equiv k+a$ and $N_+ \equiv k-a$, respectively. Out of
these there are $p_\pm \equiv \frac{q \pm a}{2}$ particles and $h_\pm
\equiv N_\pm - p_\pm = k- p_\mp$ holes. We give a few identities that
clarify the significance of these quantities:
\ba{identkaq}
N_+ + N_- &=& 2k \nn \\
p_+ + p_- &=& q \nn \\
h_+ + h_- &=& 2k-q \nn \\
p_+ - p_- &=& h_+ - h_- =\frac{1}{2} (N_+ - N_-) =a  \\
p_+ - h_+ &=& p_- - h_- = q - k \nn \\
p_- + h_+ &=& p_+ + h_- = k \nn \,\, .
\ea
Furthermore, we introduce  $h_C \equiv K-k$, which is the number of
contracted hole indices of $\V kaq$. The contracted indices never
appear with exciton operators.
Since as a consequence of~(\ref{exop}) hole indices associated with
creators (annihilators) appear in the bra (ket) of the \me , the
general structure of $\V kaq$ is~\cite{sign}:
\ba{gestVkaq}
\V kaq &=&   \es {
                   \alpha_1 \ldots \alpha_{h_C}
                 }
             \sum_{ \{ \beta_1 \ldots \beta_{h_-} \} }
             \sum_{ \{ s_1 \ldots s_{p_-}\} }
             \sum_{ \{ \gamma_1 \ldots \gamma_{h_+} \} }
             \sum_{ \{ t_1 \ldots t_{p_+}  \}  }\nn \\
&\phantom{=}& \bra{ \alpha_1 \ldots \alpha_{h_C}
                   \beta_1 \ldots \beta_{h_-}
                   t_1 \ldots t_{p_+}
                 }
              \tilde{{\sf v}}
              \ket{ \alpha_1 \ldots \alpha_{h_C}
                   \gamma_1 \ldots \gamma_{h_+}
                   s_1 \ldots s_{p_-}
                  }  \nn \\
&\times&     {\sf a}^\dagger_{t_{p_+}}
             \ldots
             {\sf a}^\dagger_{t_{1}}
             {\sf a}^\dagger_{\gamma_{1}}
             \ldots
             {\sf a}^\dagger_{\gamma_{h_+}}
             {\sf a}_{\beta_{h_-}}
             \ldots
             {\sf a}_{\beta_{1}}
             {\sf a}_{s_{1}}
             \ldots
             {\sf a}_{s_{p_-}}   \nn \\
&=&          \mes {H_C}{H_-P_-}{H_+P_+}
             \bra {H_{C}H_{-}P_+} \tilde{{\sf v}}
             \ket {H_{C}H_{+}P_-}
          \left(
             {\sf A}^{P_+}_{H_+}
          \right)^\dagger
          \left(
             {\sf A}^{P_-}_{H_-}
          \right) \,\, .
\ea
\subsection{Partial and total level densities}\label{pvz}

In the statistical model of CN processes, reaction rates are dominated
by the available phase space. This calls for a detailed knowledge of
the \dss , which have been the subject of intensive studies, see for
example~\cite{Eri,GC,Sto,GH}. In the present section, we quote the
results relevant for the sequel. The \ds \ for the $A$-body system,
\be{dichteA}
\rho ({\cal E})= \es \Sigma \delta({\cal E - E}_\Sigma) \,\, ,
\ee
is approximated by continuous expressions derived with methods of
statistical mechanics. The famous Bethe formula is
\be{Bethe}
\rho^{({\rm B})} (E) = \frac{{\rm
exp}[2({\pi^2}gE / {6} )^{1/2}]}{\sqrt{48}E}.
\ee
Here, $E$ is the {\sl excitation} energy of the system, $E \equiv
{\cal E -E}_A$ (with the ground state energy ${\cal E}_A$) and $g$
is the single body density at the Fermi edge. Blatt and
Weisskopf~\cite{BW} give the expression
\be{BW}
\rho^{({\rm BW})} (E) = C {\rm exp}[2(aE)^{1/2}]\,\, ,
\ee
which may be understood as an approximation to eq.~(\ref{Bethe})
in so far as it contains only the term varying most rapidly with energy.
Finally the approximation of constant temperature (CTA) yields a
purely exponential increase for the nuclear level density, i.e.
\be{cta}
\rho^{({\rm CTA})} (y) = \rho (y_0) {\rm e}^{(y - y_0)/T} \,\, .
\ee
Here we have called the excitation energy $y$, because we will need the
total level density in this form in section~\ref{swabs}. The nuclear
temperature is defined as
\be{temp}
 T =  \sqrt{\frac{6 y_0}{\pi^2g}} \,\, .
\ee

Gilbert and Cameron~\cite{GC} and v.~Egidy and
collaborators~\cite{TvE} have compared these expressions with
experimental data on nuclear level densities and found that the CTA
gives a good fit up to excitation energies of approximately 10 MeV,
whereas above this value the Bethe or Blatt and Weisskopf expressions
must be used.

Since we are going to work with the exciton picture, we need expressions
for the \dss \ with fixed exciton number. They will be characterized by
the number of holes and particles that occur:
\be{partzus}
  \des HP \delta(E -  E^P_H)
  \equiv \D hpE = \D {p-\Delta}pE \,\, .
\ee
Here, $E^P_H = {\cal E}^P_H - {\cal E}_A = \epsilon^P_H + {\cal E}_C
+\varepsilon_C \Delta -{\cal E}_A \equiv \epsilon^P_H +{\cal S}$. The
shift ${\cal S}$ is nonzero only if $C\neq A$.
Summation over the subspaces yields the total \ds :
\be{partvoll}
\sum_{p={\rm max}(\Delta,0)}^A \D{p-\Delta}pE
                =
                  \rho (E) \,\, .
\ee
The identity~(\ref{partvoll}) is of course independent of $\Delta$. A
different  $\Delta$ merely implies a different description, but does
not alter the physics of the level density of the $A$-body system:
\be{rhoinv}
\sum_{p={\rm max}(\Delta,0)}^A
\des {P-\Delta,}P \delta( E -  E^P_{P-\Delta})
                =
\sum_{p={\rm max}(\Delta',0)}^A
    \des {P-\Delta',}P \delta( E- E^P_{P-\Delta'})
                =
                  \rho (E) \,\, .
\ee
Since we are going to make use of this invariance property of $\rho$
in section~\ref{PA}, it is necessary to explain in detail how it is to
be understood. The sum over $p$ on the r.h.s. of eq.~(\ref{rhoinv})
may contain subspaces (e.g.   1$p$ 3$h$
configurations) that do not appear at all on the left hand side (which
could start with the subspace of 2$p$ 0$h$
configurations). And also the single exciton spaces are different: an
increase of $\Delta$ decreases the dimension of the single hole space
and enlarges that of the single particle space. If we assume a
constant spacing of the single body levels, however,
the single exciton energies $\epsilon_{r_i}$ and $\epsilon_{\alpha_i}$
in eqs.~(\ref{enerp}) and~(\ref{enerh}) that occur in the summations
over the subspace configurations are not affected by the
transformation~(\ref{rhoinv}) --- except for the highest
excited states. We can safely ignore this difference in the
energy range of interest. Explicitly, the identity~(\ref{rhoinv})
then reads
\be{rhoinvexp}
\sum_{p={\rm max}(\Delta,0)}^A
\des {P-\Delta,}P \delta( E - \epsilon^P_{P-\Delta} - {\cal S} )
                =
\sum_{p={\rm max}(\Delta',0)}^A
    \des {P-\Delta',}P \delta( E - \epsilon^P_{P- \Delta'} - {\cal S}')
              \,\, ,
\ee
where ${\cal S}' =  {\cal E}_{C'} +\varepsilon_{C'} \Delta' -{\cal
E}_A$, see below eq.~(\ref{partzus}).
Ericson~\cite{Eri} gave the
well-known approximate expression
\be{Eri}
\DE hpE      =      \frac{g[g(E- {\cal S})]^{h+p-1}}{h!p!(h+p-1)!} \,\, ,
\ee
for the partial \ds ~(\ref{partzus}). It is valid if the Pauli
principle for excitons is ignored. This corresponds to the ``dilute
gas approximation'' discussed below. Eq.~(\ref{Eri}) was lateron
improved in many respects (e.g.~\cite{Wil,Oblo}), but is widely used
because of its simplicity.

\section{Procedure}\label{proc}
\subsection{Propagation into the Subspace}\label{PS}
The strength function for transitions between states of energy $E$ and
exciton number  $N=p+h$ and states of energy $E'$ and exciton number
$N'=t+l$ is defined as
\be{locstr}
 S_{(t,p)} (E',E)
= \ves {LT}{HP}
 \B{T}{L}{\sf V} \K PH ^2 \delta(E'-E^{T}_{L})\delta(E-E^P_H).
\ee
The local average $<{\sf V}^2>$ over the squared \me s between states
with $E,N$ and $E',N'$ then is
\be{locme}
<{\sf V}^2> \D lt{E'} \D hpE =  S_{(t,p)} (E',E)  \,\, ,
\ee
where the partial  \dss \  are defined in eq.~(\ref{partzus}).
Hence, the average is defined with the delta functions of the energies
as weighting factors. However, in the present context
$\delta(E-E^P_H)$ is to be understood not as the Dirac
distribution but rather as a peaked function of suitable width.
The ``suitable width'' is large compared to the average level spacing
and small compared to intervals over which secular variations of the
level densities occur. This guarantees that \dss \ and strength
functions are smooth and can be reasonably approximated by the
expressions of subsection~\ref{pvz}.
The strength produced by the \ia \ operator $\V kaq$ is by
specialization of eq.~(\ref{locstr}) and use of eq.~(\ref{gestVkaq}):
\ba {locstrk1}
\lefteqn{ \St kaqtp (E',E)
\equiv
 \ves {LT}{HP}
 \B{T}{L} \V kaq \K PH ^2 \delta(E'-E^{T}_{L})\delta(E-E^P_H)} \\
&=&
 \ves {LT}{HP} \hspace{-0.3cm}
 \B{T}{L} \hspace{-0.15cm}
 \left[ \mes {H_C}{H_+P_+}{H_-P_-}  \hspace{-0.5cm}
     \bra {H_{C}H_-P_+} \tilde{{\sf v}}
     \ket {H_{C}H_+P_-}
       \left(
         {\sf A}^{P_+}_{H_+}
      \right)^\dagger
      \left(
         {\sf A}^{P_-}_{H_-}
      \right)
 \right] \hspace{-0.15cm}
 \K PH ^2  \delta(E'-E^{T}_{L})\delta(E-E^P_H). \nn
\ea
Note that of the seven quantities that specify the strength only six
are independent since $t=p+a$. Furthermore we observe that a nonzero
contribution comes only from the subspaces with $p \geq p_-$ and $h
\geq h_-$, because otherwise the \me \ vanishes.  We proceed to
estimate this expression in three steps.

(i) In the first step the $K$-body \me s of ${\sf V}$ are considered
to be entries of a random matrix. Invoking time reversal
invariance we postulate that ${\sf V}$ (in $K$-body space) belongs to
the Gaussian Orthogonal Ensemble (GOE). This means that the second
moments of the \me s in eq.~(\ref{opbzd}) can be expressed by a
single parameter ${v^2}$, namely (a bar over a symbol denotes
the ensemble average)
\be{ensavbp}
\overline{\bra{K_1}\tilde{{\sf v}}\ket{K_2}
 \bra{K'_1}\tilde{{\sf v}}\ket{K'_2}}
=
{v^2}[\delta_{K_1}^{ K'_1} \delta_{K_2}^{ K'_2}
     + \delta_{K_1}^{ K'_2} \delta_{K_2}^{ K'_1}]
\ee
for $K$-body configurations $K_1,K_2, K'_1,K'_2 $. Here,
$\delta^{K}_{K'}$ is a generalized Kronecker symbol~\cite{LR} with the
properties
\be{gKs}
\delta^{K}_{K'} =
\left\{ \begin{array}{r@{\quad:\quad}l}
              1  & {\rm if \ }K' {\rm \ is \ an \ even \ permutation \
of \ }K  \\
             -1 & {\rm if \ } K' {\rm \ is \ an \ odd \ permutation \
of \ }K\\
              0  & {\rm if \ two \ indices \ out \ of \ }K{\rm \ or \
out \  of \ }K'
                           {\rm \ coincide }\\
              0 & {\rm  if \ }K'{\rm \ is \
not \ a
\                            permutation \  of \ }K \,\, .
         \end{array} \right.
\ee
This allows us to estimate the strength function~(\ref{locstrk1}) by its
ensemble average, or --- by the same token --- the spectral average
$<{\sf V}^2>$ is identified  with the ensemble average.

(ii) How does the correlation rule~(\ref{ensavbp}) translate into
the exciton picture? We find the following relation:
\be{ensavexact}
 \overline{\bra{H_{C}H_{-}P_+}\tilde{{\sf v}}\ket{H_{C}H_{+}P_-}
 \bra{L_{C}L_{-}T_+}\tilde{{\sf v}}\ket{L_{C}L_{+}T_-}}
=
 {v^2}[\delta_{H_{C}H_{-}P_+}^{L_{C}L_{-}T_+}
                \delta_{H_{C}H_{+}P_-}^{L_{C}L_{+}T_-}
               +
                \delta_{H_{C}H_{-}P_+}^{L_{C}L_{+}T_-}
                \delta_{H_{C}H_{+}P_-}^{L_{C}L_{-}T_+}].
\ee
In principle, eq.~(\ref{ensavexact}) allows the simplification of the
various terms in the strength function~(\ref{locstr}). The evaluation
of the sum over the hole indices $\{H_C;H_+;H_-\}$
in eq.~(\ref{locstrk1}), however, is
complicated by the fact that it is not completely restricted in the
sense of eq.~(\ref{eingsum}). We therefore use an approximation to
eq.~(\ref{ensavexact}):
\be{ensav}
 \overline{\bra{H_{C}H_{-}P_+}\tilde{{\sf v}}\ket{H_{C}H_{+}P_-}
 \bra{L_{C}L_{-}T_+}\tilde{{\sf v}}\ket{L_{C}L_{+}T_-}}
=
 {v^2}\delta_{H_C}^{L_C}[\delta_{H_{-}P_+}^{L_{-}T_+}
                \delta_{H_{+}P_-}^{L_{+}T_-}
                +
                \delta_{H_{-}P_+}^{L_{+}T_-}
                \delta_{H_{+}P_-}^{L_{-}T_+}].
\ee
The quality of this approximation is discussed in
appendix~\ref{goecorr}, where the additional correlations resulting
from eq.~(\ref{ensavexact}) are found to be negligible. Note that
rule~(\ref{ensav}) implies that different exciton operators are
uncorrelated. Therefore the cross terms appearing in~(\ref{locstr}) do
not contribute to the strength function, which may consequently be
obtained by summing eq.~(\ref{locstrk1}) over $k,a$ and $q$.  With
the approximate correlation rule one easily arrives at:
\be{locstrk2}
 \St kaqtp(E',E)
=
 {v^2} \hspace{-0.4cm} \mes {H_C}{H_+ P_+}{H_- P_-}
             \ves {LT}{HP}
 \B{T}{L}
      \left(
         {\sf A}^{P_+}_{H_+}
      \right)^\dagger
      \left(
         {\sf A}^{P_-}_{H_-}
      \right)
  \K PH ^2
  \delta(E'-E^{T}_{L})\delta(E-E^P_H) \,\, .
\ee
The contribution of the second term on the r.h.s. of eq.~(\ref{ensav})
is restricted to the diagonal elements of the matrix and is therefore
neglected relative to the first one.
By eq.~(\ref{locstrk2}) the problem factors into two aspects: All
information that is
specific for the \ia \ is contained in  ${v^2}$. In the present paper
this factor is taken for granted. The remaining sum is common to all
\ia s of body rank $K$. It represents
the phase space aspect of the problem, the propagation of
the $K$-body force into the $A$-body space.\\
The summation over $H_C$, which appears for all operators with $k<K$
is now trivial and yields the factor ${d_h \choose h_C}$.

(iii) The sum over $\{LT;PH\}$ is rewritten in a form which one can
call the separation of actors and spectators. The configurations
$\{H_+P_+\}$ and $\{H_- P_-\}$ that appear in the operators ${\sf A}$
are called actors. They must appear in $\B TL$ and $\K PH $,
respectively, if the \me \ in eq.~(\ref{locstrk2}) is to be different
from zero. The remaining configurations that are possibly present in
$\B TL$ and $\K PH $ are called spectators. As proven in
appendix~\ref{asapp} one finds
\ba {ACTSPEC}
\lefteqn{
      \ves {LT}{HP}
      \B{T}{L}
      \left(
         {\sf A}^{P_+}_{H_+}
      \right)^\dagger
      \left(
         {\sf A}^{P_-}_{H_-}
      \right)
  \K PH ^2 \hspace{0.3cm} f(L,T,H,P) \nn} \\
&=&
 \sum_{\{H-h_-,P-p_-\} \atop \neq H_-P_- H_+P_+}
  f((H-h_-,H_+),(P-p_-,P_+),(H-h_-,H_-),(P-p_-,P_-)) \,\, ,
 \ea
where $f$ is any function that is completely symmetric in the indices
contained in $L,T,H$ and $P$ (for each group separately). The sum on
the r.h.s. of eq.~(\ref{ACTSPEC}) runs over the spectators. The
notation $\{H-h_-,P-p_-\}$ means a string of $h-h_-$ indices of holes
and a string of $p-p_-$ indices of particles that observe the
restrictions of eq.~(\ref{eingsum}). In eq.~(\ref{ACTSPEC}), there is
the additional restriction $ \neq H_- P_- H_+ P_+$ which means that
none of the indices in $\{H-h_-,P-p_-\}$ is allowed to coincide with
any one of the indices in $  H_- P_- H_+ P_+$. By help of
eq.~(\ref{ACTSPEC}) one can simplify eq.~(\ref{locstrk2}) as follows
\ba{locstrk3}
\lefteqn{ \St kaqtp(E',E)
=
 {v^2}  {d_h \choose h_C} \hspace{-0.2cm}
 \ves {H_+ P_+}{H_- P_-} \hspace{-0.1cm}
 \sum_{\{H-h_-,P-p_-\} \atop \neq H_+P_+  H_-P_-} \hspace{-0.7cm}
 \delta(E'-E^{P-p_-}_{H-h_-}-\epsilon^{P_+}_{H_+})
 \delta(E-E^{P-p_-}_{H-h_-}-\epsilon^{P_-}_{H_-})} \nn \\[2em]
&=&
 {v^2} {d_h \choose h_C} \ves {H_+ P_+}{H_- P_-}
 \sum_{\{H-h_-,P-p_-\}
         \atop \neq H_+P_+ H_-P_-}
 \int {\rm d}y \,\, \delta(E' - y - \epsilon^{P_+}_{H_+})
               \delta(E- y - \epsilon^{P_-}_{H_-})
               \delta(y - \epsilon^{P-p_-}_{H-h_-} - {\cal S})  ,
\ea
where we have used $E^{P-p_-}_{H-h_-} = \epsilon^{P-p_-}_{H-h_-} -
{\cal S}$. As already mentioned, the sum over the spectators is not
independent of the sum over the actors: The spectators are not allowed
to occupy the exciton states of the actors. This is only a weak
condition if the dimensions of the single particle and single hole
spaces are much larger than the number of excitons that occur. If the
energies $E$ and $E'$ are not too high, this will be true and one can
treat the sums in eq.~(\ref{locstrk3}) as independent. This is called
the dilute gas approximation (DGA). It allows us to express the
strength function by the convolution of partial \dss
\be {locstrk4}
 \St kaqtp (E',E)
=
 {v^2} {d_h \choose h_C}
 \int {\rm d}y\, \D {h_+}{p_+}{E'-y + {\cal S}} \D {h_-}{p_-}{E-y
 + {\cal S}} \D{h-h_-}{p-p_-}y \,\, .
\ee
As discussed in detail in appendix~\ref{goecorr}, the conditions for
the validity of the DGA and for the applicability of the approximate
correlation rule~(\ref{ensav}) are essentially the same.

\subsection{Propagation into the $A$-body space}\label{PA}
In section~\ref{PS}, the \me \ was averaged over configurations with
given exciton numbers. This result is useful if pre-equuilibrium
reactions are studied. In equilibrium CN reactions one asks for the
average $\ll {\sf V}^2 \gg $ over the full $A$-body space which is
defined as
\ba{vlocstr}
 S\,(E',E)
&\equiv&
 \ll{\sf V}^2 \gg \rho \ (E') \rho \ (E)  \\
&=&
 \sum_{t,p={\rm max}(\Delta,0)}^A
 \ves {LT}{HP}
 \B{T}{L}{\sf V} \K PH ^2 \delta(E'-E^{T}_{L})\delta(E-E^P_H) \nn
\ea
in close analogy with eqs.~(\ref{locstr}) and~(\ref{locme}).
The summation of the partial strengths up to $p,t=A$ is formally
correct although in the energy range we are interested in (and committed
to because of the DGA) by far not all subspaces come
into play. The high energy subspaces are excluded by the
multiplication with $\delta$-functions. As
discussed in the last section and in appendix~\ref{goecorr}, different
exciton operators are uncorrelated in the framework of the DGA. We
therefore obtain the strength $S(E',E)$ as a sum of the contributions
by the operators $\V kaq$:
\be{sumkaq}
S(E',E)
 =
\sum_{kaq}^{} \  \Sv kaq (E',E) \,\, ,
\ee
in obvious notation. These contributions, in turn, are easily
expressed by the strength
functions~(\ref{locstrk1}):
\be{vlocstrk1}
 \Sv kaq (E',E)
=
 \sum_{p={\rm max}(\Delta+k-\frac{q+a}{2},\frac{q-a}{2})}^{{\rm min}(A,A-a)}
 \ \St kaq{p+a}p (E',E)\, ,
\ee
where the condition $t=p+a$ has been used to evaluate one of the sums
over the subspaces. The lower limit of the remaining sum guarantees
that $p \geq p_-$, $h \geq h_-$ (see below eq.~(\ref{locstrk1})) and $p
\geq \Delta$ (see eq.~(\ref{pminush})). The upper limit ensures that
$p,t \leq A$.  We introduce the index of summation $i=p-\frac{q-a}{2}$
and the strength becomes:
\be{vlocstrk2}
 \Sv kaq (E',E)
=
 \sum_{i={\rm max}(\Delta+k-q,0)}^{A-\frac{q+|a|}{2}}
 \ \St kaq{i+p_+}{i+p_-} (E',E)  \,\, .
\ee
The second form of eq.~(\ref{locstrk3}) then yields
\ba{vlocstrk3}
     \Sv kaq (E',E)&=&  {v^2}  {d_h \choose h_C}
   \sum_{{i={\rm max}
           \atop
        (\Delta+k-q,0)}}^{A-\frac{q+|a|}{2}}
   \des{I-\Delta-k+q,}{I}
   \int {\rm d} y\, \D {h_+}{p_+}{E'-y + {\cal S}}
                    \D{h_-}{p_-}{E-y + {\cal S} } \nn \\
&\times&
   \delta(y - \epsilon^I_{I-\Delta-k+q} - {\cal S})  \,\, .
\ea
Here, the partial level densities of actors are the same as those in
eq.~(\ref{locstrk4}). In eq.~(\ref{vlocstrk3}) the expression
\be{R}
{\rm R}(y) \equiv   \sum_{p={\rm max}(\Delta',0)}^{A'}
                       \des{P-\Delta',}{P}
      \delta(y - \epsilon^P_{P-\Delta'} - {\cal S})\,\, .
\ee
appears with $A'=A-\frac{q+|a|}{2}$ and $\Delta'=\Delta+k-q \equiv \Delta
+z$. We want to compare R with the total \ds \ $\rho$ of the $A$-body
system, the definition~(\ref{partvoll}) of which is quite similar to
the expression~(\ref{R}). The comparison with $\rho (y)$ is
possible if we exploit the invariance of the nuclear level density
under shifts of the exciton vacuum. We recall eq.~(\ref{rhoinvexp}):
\be{dichtenerc'}
       \rho (y)
=
       \sum_{p={\rm max}(\Delta',0)}^{A}
       \des {P-\Delta',}P
       \delta(y - \epsilon^P_{P-\Delta'}- {\cal S}')  \,\, .
\ee
Expressions~(\ref{dichtenerc'}) and~(\ref{R}) differ in two
respects :
\begin{enumerate}
\item
    The subspaces with $p= A'+1 \ldots A$ do not appear in
    eq.~(\ref{R}).
\item
    The argument is shifted by ${\cal S}'- {\cal S}$.
\end{enumerate}
In the energy range of typical CN reactions the first point is
irrelevant so that the only remaining difference is the shift of the
argument:
\be{vgl1}
              {\rm R} (y)
   =
              \rho ( y +  {\cal E}_{C'} - {\cal E}_C
                +  \varepsilon_{C'} \Delta'
                -  \varepsilon_{C} \Delta )
\ee
In the approximation of  equidistant single body levels we find the
relation:
\be{vgl2}
              R (y)
=
              \rho \left(y   - \frac{z}{g} \left( \frac{z+1}{2}
                            +\Delta\right) \right).
\ee
At first sight one may be surprised to find that this relation depends
on $\Delta$, which {\it a priori} we may choose arbitrarily. On the
other hand, however, the choice of $\Delta$ determines the quality of
the DGA: it is best if those subspaces that contribute most to the
strength are made up of as few excitons as possible. The more states
are excluded by the Pauli--principle, the larger is the error in
eq.~(\ref{vlocstrk3}). Evaluating the integral, we choose  $\Delta=0$,
which optimizes the DGA. Hence,
\be{vlocstrk5}
             \Sv kaq (E',E)
=
            {v^2} {d_h \choose h_C}
            \int {\rm d}y\, \D {h_+}{p_+}{E'-y} \D {h_-}{p_-}{E-y}
            \rho \left( y - E_z \right) \,\, ,
\ee
with
\be{Ez}
E_z   =   \frac{z(z+1)}{2g} \,\, ,
\ee
which is the energy needed to create $z$ particles (negative $z$
corresponds to the creation of holes).

\section{Results}\label{res}
\subsection{Transition rates in the exciton model} \label{trem}
Formula~(\ref{locstrk4}) can be applied to the exciton model of
preequilibrium nuclear reactions. In this model, which was formulated
by Griffin~\cite{Gr} and has later been refined by several
authors~\cite{Cl,Ri,Bl,Wei}, an important concept is that of
transition rates between subspaces of different exciton number. These
appear in the Master equation for the time dependence of the
system. Transitions are assumed to be caused by the residual \ia \ of
a strong two-body force ${\sf V}$. The rate for going from the
subspace ${\cal U}_p$ to ${\cal U}_{p+a}$ is given by
\be{transratedef}
\lambda_{p \rightarrow p+a}  (E)     =   \frac{2\pi}{\hbar} <{\sf V}^2>
                                                     \D{h+a}{p+a}E
\ee
and is related to the strength calculated in section~\ref{PS}
according to
\be{transratestr}
\lambda_{p \rightarrow p+a} (E){\D hpE}    =   \frac{2\pi}{\hbar}
                             \sum_q \ \St 2aq{p+a}p (E,E) \,\, .
\ee
Note that the residual \ia \ consists by definition of the operators
with exciton rank $k=2$ that appear if ${\sf V}$ is expressed in the
exciton picture using the Hartree-Fock single exciton configurations.
Invoking the Ericson densities~(\ref{Eri}) and $\Delta=0$, the
convolution~(\ref{locstrk4}) can easily be evaluated and leads to the
transition rate
\be{transratprop}
   \lambda_{p \rightarrow p+a} (E) \ =  \ \frac{2\pi}{\hbar}
    v^{2} \, \sum_q
          {}_{\phantom{N}}^{\phantom{()}2\!}\rho^{(a,q)}_{h,p}(E)\,\, ,
\ee
with the density of final states
\be{findens}
   {}_{\phantom{N}}^{\phantom{()}k\!}\rho^{(a,q)}_{h,p} (E)
   = \ {d_h \choose h_C} {p \choose p_-}{h \choose h_-}
  {N+k+a-2 \choose N-1}^{-1} {2k-2 \choose k+ a -1} \
    \rho_{h_{_{+}},p_{_{+}}} (E) \,\, ,
\ee
where $N=p+h$, see above eq.~(\ref{locstr}).  Because of the last
binomial only operators with $k>|a|$ contribute. The final state
densities that appear here in the context of propagation of the
defining GOE \me s have been obtained earlier for $k=2$ by
combinatorial arguments on the states accessible in two-body
collisions~\cite{Wi,Ob,Oblo}. Originally, $v^2$ was a fit
parameter. It is identified here as the average square of the \as \
$K$-body \me .

\subsection{Spreading width in $A$-body space} \label{swabs}
Formula~(\ref{vlocstrk5}) can be evaluated by inserting Ericson's
expressions for the partial and the CTA for the total \ds , see
eqs.~(\ref{Eri}) and~(\ref{cta}) . The operator $\V kaq$ then yields
the strength:
\be{vlocstrk6}
       \Sv kaq (E)
=
       {v^2}  {d_h \choose h_C} {2k-2 \choose k+a-1} \
       \frac{g(gT)^{2k-1}  }{p_+! h_+! p_-! h_-!} \hspace{0.5cm}
       {\rm exp}\left(-\frac{z(z+1)}{2gT}\right) \hspace{0.5em}
       \rho (E).
\ee
We have chosen $y_0 = E$ in eq.~(\ref{cta}).
As mentioned in section~\ref{pvz}, below $E \approx 10$ MeV one should use
the temperature $T$ tabulated in ref.~\cite{TvE}. For $E>10$ MeV, the
temperature should be determined from eq.~(\ref{temp}), again with
$y_0=E$~\cite{ehat}.
In experiment, of course, the effect of the \ia \ ${\sf V}$ as a whole
is measured.  The \sw ~(\ref{spreadw}) is obtained by summing
eq.~(\ref{vlocstrk6})  over $k,a$ and $q$ after
dividing through $\rho (E)$. This gives:
\be{gammaK}
\swd (E) =       2 \pi {v^2} \sum_{kaq}
                 {d_h \choose h_C} {2k-2 \choose k+a-1} \
                 \frac{g(gT)^{2k-1}  }{p_+! h_+! p_-! h_-!} \hspace{0.3cm}
                 {\rm exp}\left(-\frac{z(z+1)}{2gT}\right) \,\, .
            \ee
Usually a two-body ansatz is made for the \ia s between nucleons. We
therefore explicitly carry out the summation in eq.~(\ref{gammaK}) for
this case. The factor exp($-z(z+1)/2gT$) by which the total spectator
density deviates from the nuclear level density ---  close to unity for
typical values of $T,g$ and $E$ --- is ignored here in order
to analyze the general properties of the spreading width. One finds:
\be{Gamma2}
\swd (E) =2\pi\, {v^2} \, g^2 T [ 2d_h + 5(gT)^2 ] \,\, ,
\ee
which is a remarkably simple result. We emphasize two aspects of it:

(i) According to ref.~\cite{TvE} this is a constant as a function of
$E$ below $E\approx 10$ MeV because the nuclear temperature should
then be independent of $E$. At higher energy the leading term behaves
as $E^{3/2}$. Altogether this amounts to quite a weak
energy dependence of $\swd$ for moderate $E$ --- especially if
compared to the exponential energy dependence of $\rho$. This result
is in qualitative agreement with the systematics of the \sw s
pertaining to isospin violation~\cite{RH}. Eq.~(\ref{gammaK}) shows
that the leading energy dependence of $\swd$ will be $E^{(2K-1)/2}$ if
the body rank of ${\sf V}$ is $K$ instead of two. This demonstrates
that the reason for the weak energy dependence of the experimental
$\swd$ is the two-body character of the symmetry breaking \ia : With
increasing excitation energy the complexity (in terms of excitons) of
the states increases. This decreases the fraction of states that can
be connected by an \ia \ of low rank, hence the local average square
$\ll{\sf V}^2 \gg$ decreases. Consequently, the product of $\ll{\sf
V}^2 \gg$ and $\rho$ varies slowly.

(ii) The contributions ${^k\swd}$ of the exciton potential (k=1) and
the exciton scattering (k=2) to the \sw \ are given by the first and
second term on the r.h.s. of eq.~(\ref{Gamma2}), i. e. by the terms
proprotional to $T$ and $T^3$ respectively.  Inserting the typical
value of $T=0.5$ MeV for $E\lesssim 10$ MeV~\cite{TvE} and $g \approx
{A}\ (13 \ {\rm MeV})^{-1}$ as well as $d_h=A$ (which optimizes the
DGA), one finds
\be{1Gammazu2Gamma}
\frac{^1\swd}{^2\swd} \approx \left\{ \begin{array}{r@{\quad:\quad}l}
                    1 & E \le 10 \, {\rm MeV}  \\[1cm]
                    10\,{\rm MeV}/E  & E>10\,{\rm MeV}.
                           \end{array} \right.
\ee
Thus the rank one and rank two exciton \ia s contribute about equally
strongly to $\swd $  in the energy range of typical CN reactions.

\subsection{Discussion}

Within the framework of the statistical model the \sw \ due to an
arbitrary $K$-body force has been calculated for a compound nuclear
reaction of an $A$-body nucleus. The \sw \ is the
adequate measure for symmetry nonconservation in complex many-body
systems~\cite{RH,Boo}. It measures the extent to which the states are
smeared out due to the presence of the symmetry breaking force. It
also appears as the damping width of a simple configuration --- such
as an isobaric analog state or a giant resonance --- into the complex
compound nuclear configurations.  Definite numbers for a particular
\ia \ can be given if eq.~(\ref{gammaK}) is complemented by the
calculation of the spectral average of the \ia \ in $K$-body
space. Obvious applications are the electromagnetic and weak forces,
responsible for the breaking of isospin and parity,
respectively. Numerical studies in that direction are currently under
progress. For
the case of isospin breaking a weak dependence of the \sw \ on energy
and mass number is known~\cite{RH}. In the present study the energy or
temperature dependence has been related to the body rank of the
underlying \ia . Since ${v^2}$ depends on $A$, eqs.~(\ref{gammaK})
and~(\ref{Gamma2}) do not exhibit the complete $A$-dependence of the
\sw , which is expected to be weak.  The variation of $\swd$ with
temperature can be compared with the results of other
authors. Assuming that the spread of a one-exciton configuration (a
1$p$ 0$h$ state) is proportional to the square of the excitation
energy~\cite{BBB,MS}, Lauritzen {\it et al.}~\cite{LDB} have concluded
a $T^3$-dependence for the spreading width in general. Since the decay
of the one-exciton configuration is caused by $\V 213$, we indeed find
from eq.~(\ref{findens}) its damping to be proportional to $E^2$ and
eq.~(\ref{Gamma2}) indeed indicates a $T^3$-dependence of the \sw \ in
$A$-body space (for transitions caused by the strong \ia \ only).
Studying the coupling of surface modes to single particle motion,
Esbensen and Bertsch~\cite{EB} found the ``elementary damping'' to be
proportional to $E$, which in the framework of Lauritzen {\it et.~al}
corresponds to a $T^2$-dependence of the spreading width. There is
nothing analogous in the present results since we did not consider
collective motion in the present paper.  De Blasio {\it
et.~al}~\cite{DeB} report the damping of giant resonances to be
independent of the nuclear temperature. Again the reason is that the
statistical damping of the present study is different from the damping
mechanism considered there. This is also true for the study of
spreading properties of isobaric analogue and Gamov-Teller resonances
by Col\`{o} {\it et al.}~\cite{CGBB}. This paper indicates, however,
that the statistical damping must be taken into account in order to
explain the widths of the resonances.

Frequently the treatment of parity violation in CN reactions is
restricted to the $k=1$ part of the weak \ia
{}~\cite{Ade,Koo,Aue,Bowm,Fla}. Eq.~(\ref{1Gammazu2Gamma}) indicates,
however, that the contribution of the operators with exciton rank
two should be included in the calculation of a local mean square \me .

Finally we point out the limitations of our method. The Hartree Fock
method and the particle hole formalism rely on a basis of product
states. Such an independent particle model can not describe collective
phenomena in nuclei. The present results therefore must be
modified if applied to reactions that involve collective excitations.
A second
problem are the effects of the symmetries {\sl respected} by the \ia \
under study. These effects will of course manifest themselves
automatically when the spectral average in the $K-$body space is
calculated.  The use of partial level densities for the actors that
contain all states at a given energy irrespective of further quantum
numbers, however, neglects possible {\sl local} effects of the
respected symmetries.  We give an example to illustrate this
complication. Consider the weak potential ($k=1$) in a nucleus. This
operator connects many-body states that differ by only one single-body
configuration. Parity violation and conservation of total angular
momentum demand that the single-body states differ in $l$ but not in
$j$. Since single-body states with $\Delta l =\pm 1$ and $\Delta j=0$
only exist in different shells, they are separated by a relatively
large energy interval. Consequently, the present local average
suppresses these contributions to the \sw . The two-exciton part of
the weak \ia \ on the other hand is not subject to this ``local
selection rule'' because it simultaneously changes two single-ecxiton
configurations. This fact was first pointed out by Lewenkopf and
Weidenm{\"u}ller~\cite{LW}. They estimated that the $k=2$ part of the
weak interaction dominates the local mean square \me . In
section~\ref{swabs} it was found that potential and scattering
contribute about equally to the \sw \ without taking the local effects
of the symmetries into account. Sufficiently elaborate expressions for
the single-exciton level densities of the actors before and after the
\ia \ would only overlap in a small energy range, and consequently the
convolution of $\D {h_+}{p_+}{E-y} \D {h_-}{p_-}{E-y}$ with $\rho (y
- E_z)$ would result in a smaller contribution of the potential to the
strength.

\section*{Acknowledgements}
M.G. wishes to thank Simon Kalvoda for stimulating discussions and
the {\it Studienstiftung des deutschen Volkes} for support.

\appendix

\section{GOE correlations in the exciton picture}\label{goecorr}

In this appendix, the differences between the
exact~(\ref{ensavexact}) and approximate~(\ref{ensav}) ensemble average
are discussed. We find that eq.~(\ref{ensavexact}) leads to
two types of additional correlation that do not appear in its
approximation. Considering two examples, it will also be found,
however, that these additional terms are of the same order of
magnitude as those
neglected by invoking the dilute gas approximation. Since the central
formulae of section~\ref{PS} and~\ref{PA} rely on the DGA, it is not
necessary and indeed would be inconsistent to take these quantities
into account in the evaluation of the strength function.

The correlation coefficient for two operators is
\ba{K0}
\lefteqn{ {\cal C}_{kk'} \equiv \B{T}{L} \V kaq \K PH
            \B{T}{L} \V {k'}{a'}{q'} \K PH
       = \sum_{
                 \{ {H_C};{H_-P_-};{H_+P_+} \}
                        \atop
                 \{ {L_C};{L_-T_-};{L_+T_+} \}
                }}\\
&\phantom{\times}&               \bra {H_{C}H_-P_+} \tilde{{\sf v}}
               \ket {H_{C}H_+P_-}
               \bra {L_{C}L_-T_+} \tilde{{\sf v}}
               \ket {L_{C}L_+T_-}\nn
           \B{T}{L}
              \left(
                    {\sf A}^{P_+}_{H_+}
              \right)^\dagger
              \left(
                     {\sf A}^{P_-}_{H_-}
              \right)
               \K PH
               \B{T}{L}
               \left(
                     {\sf A}^{T_+}_{L_+}
               \right)^\dagger
               \left(
                     {\sf A}^{T_-}_{L_-}
               \right)
               \K PH \, .
\ea
Estimating this expression by its ensemble average, we obtain:
\be{K1}
{\cal C}_{kk'} = v^2
\sum_{
                 \{ {H_C};{H_-P_-};{H_+P_+} \}
                        \atop
                 \{ {L_C};{L_-T_-};{L_+T_+} \}
                }
             \delta_{H_{C}H_{-}P_+}^{L_{C}L_{-}T_+}
                \delta_{H_{C}H_{+}P_-}^{L_{C}L_{+}T_-}
        \B{T}{L}
              \left(
                    {\sf A}^{P_+}_{H_+}
              \right)^\dagger
              \left(
                     {\sf A}^{P_-}_{H_-}
              \right)
               \K PH
               \B{T}{L}
               \left(
                     {\sf A}^{T_+}_{L_+}
               \right)^\dagger
               \left(
                     {\sf A}^{T_-}_{L_-}
               \right)
               \K PH \,\, .
\ee
Since hole and particle configurations by definition do not
intersect, the Kronecker symbol implies $p_+ =t_+$ and $ p_- = t_-$.
Addition and subtraction of these equations yields $q=q'$ and
$a=a'$. In other words: two operators that differ only
by their rank are correlated. This is the first difference to the ensemble
average~(\ref{ensav}) used in section~\ref{PS}, were it was found that
different operators are uncorrelated. The second type of additional
correlation we find in eq.~(\ref{K1}) results from the intricate
restriction pattern in the sum over hole configurations: The
restriction only applies to the indices {\sl within} the groups $L_C$
and $L_-$. The condition for nonzero correlations, however, is that
the set $L_C L_-$ of $l_C+l_-$ indices coincide with the set $H_C H_-$
of $h_C+h_-$ indices. We adress these two cases one after the other.
First, let $k'$ be equal to
$k+\xi$ with $\xi>0$. The sum over $T_+,T_-$ may now readily be
carried out:
\be{K2}
{\cal C}_{k,k+\xi} = v^2
\sum_{
                 \{ {H_C};{H_-P_-};{H_+P_+} \}
                        \atop
                 \{ {L_C};{L_-};{L_+} \}
                }
             \delta_{H_{C}H_{-}}^{L_{C}L_{-}}
             \delta_{H_{C}H_{+}}^{L_{C}L_{+}}
        \B{T}{L}
              \left(
                    {\sf A}^{P_+}_{H_+}
              \right)^\dagger
              \left(
                     {\sf A}^{P_-}_{H_-}
              \right)
               \K PH
               \B{T}{L}
               \left(
                     {\sf A}^{P_+}_{L_+}
               \right)^\dagger
               \left(
                     {\sf A}^{P_-}_{L_-}
               \right)
               \K PH \,\, .
\ee
This expression may be simplified rather easily if we consider the
special case of $l_C=0$. Then $H_C=\Xi$ and we obtain
\be{K3}
{\cal C}_{k,k+\xi} = v^2
\sum_{  \{ {\Xi};{H_-P_-};{H_+P_+} \} }
   \B{T}{L}
              \left(
                    {\sf A}^{P_+}_{H_+}
              \right)^\dagger
              \left(
                     {\sf A}^{P_-}_{H_-}
              \right)
               \K PH
               \B{T}{L}
               \left(
                     {\sf A}^{P_+}_{\Xi H_+}
               \right)^\dagger
               \left(
                     {\sf A}^{P_-}_{\Xi H_-}
               \right)
               \K PH \,\, .
\ee
For further simplification of this expression let us introduce the
product of hole number operators
\be{lxi}
({\sf L}_\Xi) \equiv {\sf a}^\dagger_{\beta_1} {\sf a}_{\beta_1}
               \ldots{\sf a}^\dagger_{\beta_\xi} {\sf a}_{\beta_\xi}
\ee
for $\Xi = (\beta_1 \ldots \beta_\xi)$. This operator has the property
\be{lxieigensch}
\es \Xi ({\sf L}_\Xi) \K PH = {h \choose \xi} \K PH \,\, ,
\ee
where  a term containing the factorial of a negative number is
understood to be zero. Note that
\be{Kid}
              |\B{T}{L}
               \left(
                     {\sf A}^{P_+}_{\Xi H_+}
               \right)^\dagger
               \left(
                     {\sf A}^{P_-}_{\Xi H_-}
               \right)
               \K PH |
=
              | \B{T}{L}
              \left(
                    {\sf A}^{P_+}_{H_+}
              \right)^\dagger
              {\sf L}_\Xi
              \left(
                     {\sf A}^{P_-}_{H_-}
              \right)
               \K PH | \,\, .
\ee
One therefore obtains
\be{K4}
|{\cal C}_{k,k+\xi}| =  {v^2}
              { h-h_- \choose \xi }
              \ves {H_- P_-}{H_+ P_+}
              \B{T}{L}
              \left(
                    {\sf A}^{P_+}_{H_+}
              \right)^\dagger
              \left(
                     {\sf A}^{P_-}_{H_-}
              \right)
               \K PH ^2 \,\, .
\ee
The quantity $h-h_-$ is the number of hole spectators, cf. section~\ref{PS}.
For the correlation coefficient of an operator with itself, we find
\be{Q}
{\cal C}_{kk}
=
              {v^2}
              { d_h \choose \xi }
              \ves {H_- P_-}{H_+ P_+}
              \B{T}{L}
              \left(
                    {\sf A}^{P_+}_{H_+}
              \right)^\dagger
              \left(
                     {\sf A}^{P_-}_{H_-}
              \right)
               \K PH ^2 \,\, ,
\ee
so that
\be{KQ1}
             \frac{|{\cal C}_{k,k+\xi}|}{{\cal C}_{kk}}
=            \frac{(h-h_-)! (d_h - \xi)!}
                  {(h-h_- - \xi)!d_h !} \,\, .
\ee
Under the assumption that $K \ll A = d_h$, which implies $\xi \ll
d_h$, this yields
\be{KQ2}
               \frac{|{\cal C}_{k,k+\xi}|}{{\cal C}_{kk}}
\lesssim     \left( \frac{h - h_-}{  d_h  } \right)^\xi \,\, .
\ee
This type of correlation may therefore be neglected if the number of
spectators (in hole space) is small compared to the dimension $d_h$ of
the single hole space. This is certainly true if $h \ll d_h$, which
is the condition for the DGA introduced in section~\ref{PS}.

The second type of additional correlation appearing in eq.~(\ref{K1})
shall be illustrated using the operator $\V 100$ of a two body \ia \
(cf.~Tab.~I). We find for its correlation coefficient:
\ba{100-1}
{\cal C}&=&v^2 \sum_{\beta \gamma \delta \atop \beta' \gamma' \delta'}
         \delta^{\beta \gamma}_{ \beta' \gamma'}
         \delta^{\beta \delta}_{ \beta' \delta'}
         \B TL {\sf a}^\dagger_\delta {\sf a}_\gamma \K PH
         \B TL  {\sf a}^\dagger_{\delta'} {\sf a}_{\gamma'} \K PH \nn \\
&=&
v^2   \sum_{\gamma \delta}
         \B TL {\sf a}^\dagger_\delta {\sf a}_\gamma \K PH
  \left(   \B TL {\sf a}^\dagger_\delta {\sf a}_\gamma \K PH d_h
        +  \Bracket TLPH h
        -2 \B TL {\sf a}^\dagger_\delta {\sf a}_\gamma \K PH
  \right) \,\, .
\ea
In the brackets, the first term is of the type discussed in
sections~\ref{PS} and~\ref{PA}. The second term is suppressed
relative to the first
one by a factor $h/d_h$ and by the fact that it contributes only to
diagonal elements.
The third term is suppressed by a factor $2/d_h$. Obviously, it is
justified to neglect the second and third term as long as the DGA
makes sense. Since out of the nine operators of a two-body \ia \
which actually contribute to the strength function $\V 100$ is the
only one that shows correlations of this type, eq.~(\ref{ensav})
seems to be a very good approximation.

\section{Proof of equation~(\ref{ACTSPEC})}{\label{asapp}}

In eq.~(\ref{locstrk2}), consider the special case of $p_+ = h_+ =h_-
= 0$:
\be{actspec1}
 {\cal M}
\equiv
 \es P \B{T}{H}
      \left(
         {\sf A}^{P_-}_0
      \right)
  \K PH ^2  f(P) \,\, .
\ee
Let $P,P_-$ be
\be{PP-}
P=(r_1 \ldots r_p) \hspace{1cm} {\rm and}
\hspace{1cm} P_-=(s_1 \ldots s_{p_-})
\ee
and $f$ a function that is completely symmetric in the arguments $r_1
\ldots r_p$. The restricted sum of eq.~(\ref{actspec1}) can be written
as the unrestricted sum
\be{actspec2}
 {\cal M}
=
  \frac{1}{p!}
  \sum_{ P }
  \B{T}{H}
    {\sf a}_{s_1} \ldots {\sf a}_{s_{p_-}}
    {\sf a}^\dagger_{r_p} \ldots {\sf a}^\dagger_{r_1}
  \K 0H^2 \hspace{0.3cm} f(P) \,\, .
\ee
The \me s vanish unless the $s_1 \ldots s_{p_-}$ all appear in $r_1
\ldots r_p$.  Consider a term that satisfies this condition. There are
${p \choose p_-} $ ways in which the indices $r$ that agree with one
of $s_1 \ldots s_{p_-}$ can be distributed over the postions $1\ldots
p$. Therefore, imposing the requirement that the first $p_-$ indices
$r_1 \ldots r_{p_-}$ should agree with $s_1 \ldots s_{p_-}$ (up to a
permutation) one obtains
\ba{actspec3}
 {\cal M}
&=&
 \frac{1}{p!} {p \choose p_-}
 \sum_{t_1 \ldots t_{p-p_-} \atop \neq s_1 \ldots s_{p_-}}
 \sum_{r_1 \ldots r_{p_-}}
  \B{T}{H}
    {\sf a}_{s_1} \ldots {\sf a}_{s_{p_-}}
    {\sf a}^\dagger_{ t_{p-p_-}} \ldots {\sf a}^\dagger_{t_1}
    {\sf a}^\dagger_{r_{p_-}} \ldots {\sf a}^\dagger_{r_1}
  \K 0H^2 \hspace{0.3cm} f(t_1 \ldots t_{p-p_-}, P_- )\nn \\
&=&
 \frac{1}{p!} {p \choose p_-} (p-p_-)!p_-!
 \sum_{{\{P-p_-\} \atop \neq P_-}}
  \B{T}{H}
    {\sf a}^\dagger_{ t_{p-p_-}} \ldots {\sf a}^\dagger_{t_1}
  \K 0H^2 \hspace{0.3cm}  f(P-p_-,P_-)\,\, .
\ea
Here, the restriction $t_1 \ldots t_{p-p_-}  \neq s_1 \ldots s_{p_-}$
means that none of the indices $t_i$ ,$i=1 \ldots p-p_-$, is allowed to
coincide with any one of the indices $s_k$, $k=1 \ldots p_-$. The short
hand notation $\{P-p_-\} \neq P_-$ means the
same. Eq.~(\ref{actspec3}) is obviously the same as
\be {actspec6}
 \sum_{{\{P-p_-\} \atop \neq P_-}}
 \Bracket{T}{H}  {P-p_-}H ^2 \hspace{0.3cm} f(P-p_-,P_-)
= \sum_{{\{P-p_-\} \atop \neq P_-}}
 \delta_{P-p_-}^{T}   \hspace{0.3cm} f(P-p_-,P_-) \,\, ,
\ee
which is easily generalized to eq.~(\ref{ACTSPEC}).

\begin{figure}[hc]
              \caption[Two-body interaction in the exciton picture]
          {
           Two-body interaction in the exciton
           representation: Diagrams with $k=1,2$. \\
           }
  \label{2biaep}
\end{figure}


\begin{references}
\bibitem{time}Usually the holes are defined as the time reversed
                 state of the corresponding body states and phases are
                 introduced in order to ensure the desired
                 transformation properties. We omit these complications
                 since they do not affect our arguments.
\bibitem{sign}We do not give the sign of the operator,
                since it is irrelevant for our purposes.
\bibitem{ehat}In order to better respect the variation of $T$ with
                 energy,
                 one could choose $y_0$ as the maximum of the
                 integrand in eq.~(\ref{vlocstrk5}). The maximum is,
                 however, close to the excitation energy $E$.
\end{references}
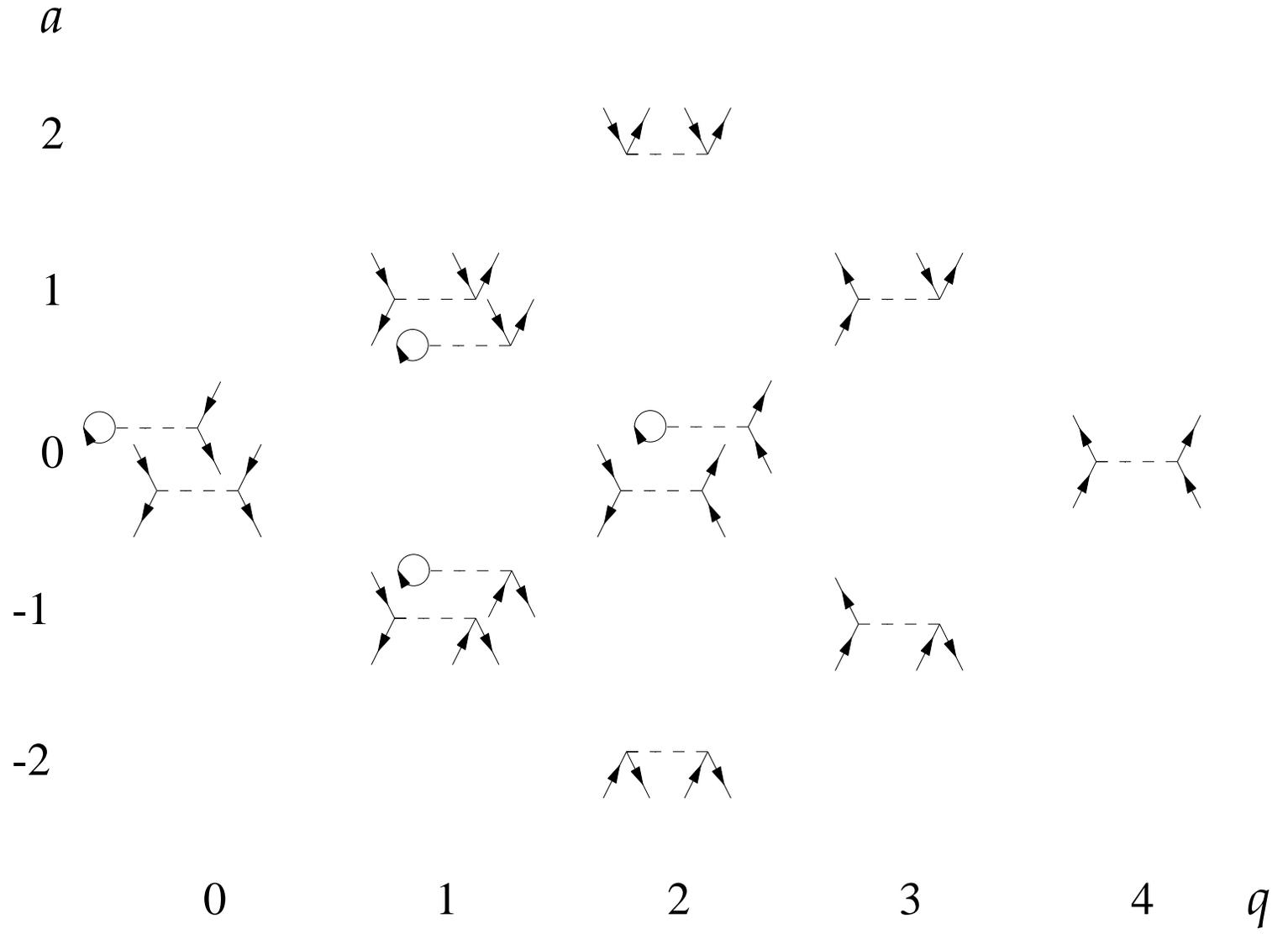
\end{document}